\documentclass[twocolumn,aps,prl,superscriptaddress,floatfix]{revtex4}
\usepackage[T1]{fontenc}
\usepackage[utf8]{inputenc}
\usepackage[english]{babel}
\usepackage{color}
\usepackage{float}
\usepackage{braket}
\usepackage{amsmath}
\usepackage{amstext}
\usepackage{amssymb}
\usepackage{graphicx}
\usepackage{bbold}
\usepackage[unicode=true,pdfusetitle,
bookmarks=true,bookmarksnumbered=false,bookmarksopen=false,
breaklinks=false,pdfborder={0 0 1},backref=false,colorlinks=false]
{hyperref}
\hypersetup{
	colorlinks,linkcolor=blue,citecolor=blue,urlcolor=blue}
\usepackage{soul}
\makeatletter
\date{}

\makeatother
\setul{0.5ex}{0.3ex}
\definecolor{Blue}{rgb}{0,0.0,1}
\setulcolor{Blue}
\usepackage{babel}

\begin{document}

\author{Luis M. Canonico}
\affiliation{Instituto de F\'\i sica, Universidade Federal Fluminense, 24210-346 Niter\'oi RJ, Brazil}
\author{Tarik P. Cysne}
\affiliation{Instituto de F\'\i sica, Universidade Federal Fluminense, 24210-346 Niter\'oi RJ, Brazil} 
\author{Alejandro Molina-Sanchez}
\affiliation{International Iberian Nanotechnology Laboratory (INL), Avda. Mestre Jos\'e Veiga s/n, 4715-330 Braga, Portugal}
\author{R. B. Muniz}
\affiliation{Instituto de F\'\i sica, Universidade Federal Fluminense, 24210-346 Niter\'oi RJ, Brazil}

\author{Tatiana G. Rappoport}
\affiliation{Instituto de F\'\i sica, Universidade Federal do Rio de Janeiro, Caixa
	Postal 68528, 21941-972 Rio de Janeiro RJ, Brazil}
\affiliation{Department of Physics and Center of Physics, University of Minho, 4710-057, Braga, Portugal}

\title{ Orbital Hall Insulating Phase in Transition Metal Dichalcogenide Monolayers }

\begin{abstract}
	We show that H-phase transition metal dichalcogenides (TMDs) monolayers such as MoS$_2$ and WSe$_2$, are orbital Hall insulators. They present very large orbital Hall conductivity plateaus in their semiconducting gap, where the spin Hall conductivity vanishes. Our results open the possibility of using TMDs for orbital current injection and orbital torque transfers that surpass their spin-counterparts in spin-orbitronics devices. The orbital Hall effect (OHE) in TMD monolayers occurs even in the absence of spin-orbit coupling. It can be linked to exotic momentum-space Dresselhaus-like orbital textures, analogous to the spin-momentum locking in 2D Dirac fermions that arise from a combination of orbital attributes and lattice symmetry. \end{abstract}
\maketitle
The flexibility to combine atom-thick layers with different characteristics in novel quantum metamaterials makes two-dimensional (2D) systems interesting platforms for spintronics \cite{spintronics2Dmaterials1, spintronics2Dmaterials2}.  In recent years, innovative routes to generate and manipulate spin currents in 2D materials where spin-orbit coupling mediates the conversion between charge and spin currents have been discovered, such as the spin Hall effect (SHE), Rashba-Edelstein effect (REE) \cite{Offidani_PRL_2017} and all-optical spin injection \cite{Avsar_ACS_2017}. Orbital angular momentum can be manipulated like spin and be relevant in many materials, even in the absence of strong spin-orbit coupling (SOC).  Orbitronics, an analog of spintronics that operates with the electronic orbital angular momentum degrees of freedom, although embryonic, is sparking renewed interests. An increasing number of effects are being predicted, such as the orbital-momentum locking \cite{OrbitalTextureBorophene}, orbital torque \cite{Orbital-torque}, orbital Rashba effect \cite{Orbital-Rashba, Orbital-Rashba_PRB_2013, Orbital-Rashba_PRB_2012}, orbital Edelstein effect \cite{Orbital-Edelstein, OrbitalEdelstein_Arxiv_2019} and orbital Hall effect (OHE) \cite{Orbitronics-Bernevig, Inoue-OHEdTransitionMetal}, an orbital analog of the SHE.  

The OHE, similarly to the SHE, refers to the creation of a transverse flow of orbital angular momentum (OAM)  that is induced by a longitudinally applied electric field~\cite{Orbitronics-Bernevig}. It has been explored mostly in three dimensional metallic systems, where it can be quite strong~\cite{Inoue-OHEdTransitionMetal,Orbitals-Frustration-Inoue,OrbitalGiantSHE-Inoue,Go-OHETexture}.  However, the OHE does not necessarily rely on strong SOC. It can be linked to orbital textures~\cite{Go-OHETexture} where the OAM is locked to the carrier momentum (similar to the spin-momentum locking observed in systems that present REE), and can be relevant in a diverse pool of materials. Recent theoretical results have predicted the existence of OHE in 2D {\it insulators}, suggesting that this effect could be found also in other elements of this class of systems \cite{us}. 

Single layers of TMDs hold great appeal for electronics, optoelectronics, and spintronics applications~\cite{TMDrev,optoTMD,fabian15, CrystalFieldGrapheneTarik,QHE-Interface-Tarik}. Two-dimensional layers of TMDs such as  MX$_2$ (M =Mo,W,and X =S,Se,Te) exhibit direct band-gap properties that are ideal for optoelectronics applications~\cite{optoTMD}. Their lack of inversion symmetry, combined with strong SOC, causes a sizeable spin splitting at the valence band edges, enabling spin- and valley-selective light absorption. These characteristics provide all-optical methods for manipulation of internal degrees of freedoms, enabling, for example, all-optical spin-injection.  Although the OAM is present in TMDs and coupled to the valley and spin degrees of freedom, the possibility of manipulating it in TMDs monolayers for spin-orbitronics applications is just beginning to be noticed~\cite{Xiao_PRL_2012, LightMatTMD}. 

Only a few studies of OHE were performed in 2D materials, partially due to their usually small orbital conductivity values \cite{GraphaneOHE, Mele_PRL_2019, us}. The search for 2D materials with a robust orbital signal is fundamental for possible developments in orbitronics. Here, we shall investigate the orbital Hall conductivity of transition metal dichalcogenide (TMD) monolayers in the H structural phase. For this purpose, we first consider a simplified tight-binding (TB) model involving only three relevant $d$ orbitals of the transition metal atoms. We then extend our analysis to a more involved multiband TB model that includes the $s$ and $p$ orbitals of the chalcogen atoms, as well as the $d$ orbitals of the TM atoms. The hopping integrals and onsite energies are obtained from \textit{ab initio} calculations~\cite{Pizzi2019}. In both cases, we demonstrate that TMDs host very robust orbital Hall currents in their insulating gap, even in the absence of spin-orbit coupling. The orbital Hall conductivity can be significantly larger than the spin Hall one and exhibits a relatively large plateau inside the TMD electronic energy gap, where the SHE is absent. 

The atomic environment of the transition metal atoms together with the interaction with the chalcogen atoms lead to a large crystal field splitting. The resulting band edges are well reproduced by a TB model comprising three atomic $d$ orbitals only arranged in a triangular lattice. This 3-bands model captures the essential features of the energy spectrum near valleys of most single layer TMD-H ~\cite{ThreebandXiao}. The TB Hamiltonian is given by
\begin{equation}
{\cal H}_0=\sum_{\langle i j\rangle} \sum_{\mu \nu s} t_{i j}^{\mu \nu}{d^\dagger_{i \mu s}}d_{j \nu s}+ \sum_{i \mu s}\epsilon_{i \mu} d^\dagger_{i \mu s} d_{i \mu s}+\sum_{i \mu \nu s} \mathbf{h}^z_{\mu\nu s}  d^\dagger_{i \mu s} d_{i \nu s} \label{eqn:HNN}.
\end{equation}

\noindent  Here, $i$ and $j$ denote the triangular lattice sites positioned at $\vec{R}_i$ and $\vec{R}_j$, respectively. $\langle i j \rangle$ indicates that the sum is restricted to the nearest neighbour (n.n) sites only. The operator $d^{\dagger}_{i \mu s}$ creates an electron of spin $s$ in the atomic orbitals $d_\mu$ located at site $i$, where $\mu=1,2,3$ represent the $d$ atomic orbitals $z^2$, $xy$ and $x^2+y^2$, respectively; $\epsilon_{i\mu}$ is the corresponding on-site atomic energy associated with orbital $\mu$, and $t_{i j}^{\mu \nu}$ are the transfer integrals between orbitals $\mu$ and $\nu$ centred on sites $i$ and $j$, respectively, constructed based on symmetry operations of the point group D$_{3h}$ of TMDs \cite{ThreebandXiao}. The  third term  describes an intrinsic atomic SOC where $\mathbf{h}^z_{\mu\nu s}=\lambda_I  \mathbf{L}^z_{\mu \nu} s^z_{s s}$. 
 
To calculate the orbital-Hall (OH) and spin-Hall (SH) conductivities for the TMDs make use of the Kubo-Bastin formula \cite{Bastin-FormulaConductividad}:
\begin{flalign}
&\sigma_{\alpha \beta}(\mu, T) = \frac{i\hbar}{\Omega}\int_{-\infty}^{+\infty}dE f(E; \mu, T) \nonumber\\ &\times Tr \langle j_{\alpha}\delta(E-\mathcal{H})j_{\beta}\frac{dG^{+}}{dE} - j_{\alpha}\frac{dG^{-}}{dE}j_{\beta}\delta(E-\mathcal{H})\rangle .
\label{eqn: KB}
\end{flalign}
Here, $\Omega$ represents the area of the sample, $f(E; \mu, T)$ is the Fermi-Dirac distribution for energy $E$, chemical potential $\mu$ and temperature $T$. $G^{+}(G^{-})$ symbolises the advanced(retarded) one-electron Green function. To calculate the SH conductivity $\sigma_{SH}^z $, we take $j_{\alpha}$ as the current-density operator component along the applied electric field direction $\hat{x}$, $j_{\alpha} \equiv j_{x} = \frac{ie}{\hbar}\left[x,\mathcal{H}\right]$, and $j_{\beta}$ as the transverse spin current-density operator component $j_\beta \equiv j_{y}^s = \frac{1}{2}\left\lbrace s_{z},v_{y}\right\rbrace$ where $s_{z}$ is the spin-Pauli's matrix and $v_{y}$ is the $y$ transverse component of the velocity operator. For calculating the OH conductivity $\sigma_{OH}^z$, we consider  $j_{\alpha} \equiv j_{x}$ and $j_{\beta}$ as the orbital transverse current density operator component $j_\beta \equiv j_{y}^{\ell_z} = \frac{1}{2}\left\lbrace \ell_{z},v_{y}\right\rbrace$, where $\ell_{z}$ is the $z$-component of the {\it atomic} angular momentum operator. 
Eq. (\ref{eqn: KB}) is equivalent to the Kubo formula written in terms of spin and orbital Berry curvatures \cite{Go-OHETexture, Mele_PRL_2019,SuplementaryMaterial}.

Our transport calculations are performed in real-space with the use of a modified version of the quantum transport software KITE~\cite{KITEzenodo} based on Chebyshev polynomial expansions~\cite{WeisseKPM}. This method is highly efficient for computation of Hall responses in 2D systems~\cite{Jose-TatianaKPM,AiresCriticalDelocalizatoin,PRLPxPy-Nosotros,Jose-tatiana,JoseTMDGraphene}. Our simulations were performed for systems with $2\times 512\times 512$ unit cells, including up to 1024 moments in the Chebyshev polynomial expansion
\begin{figure}[h!]
	\includegraphics[width=0.95\linewidth,clip]{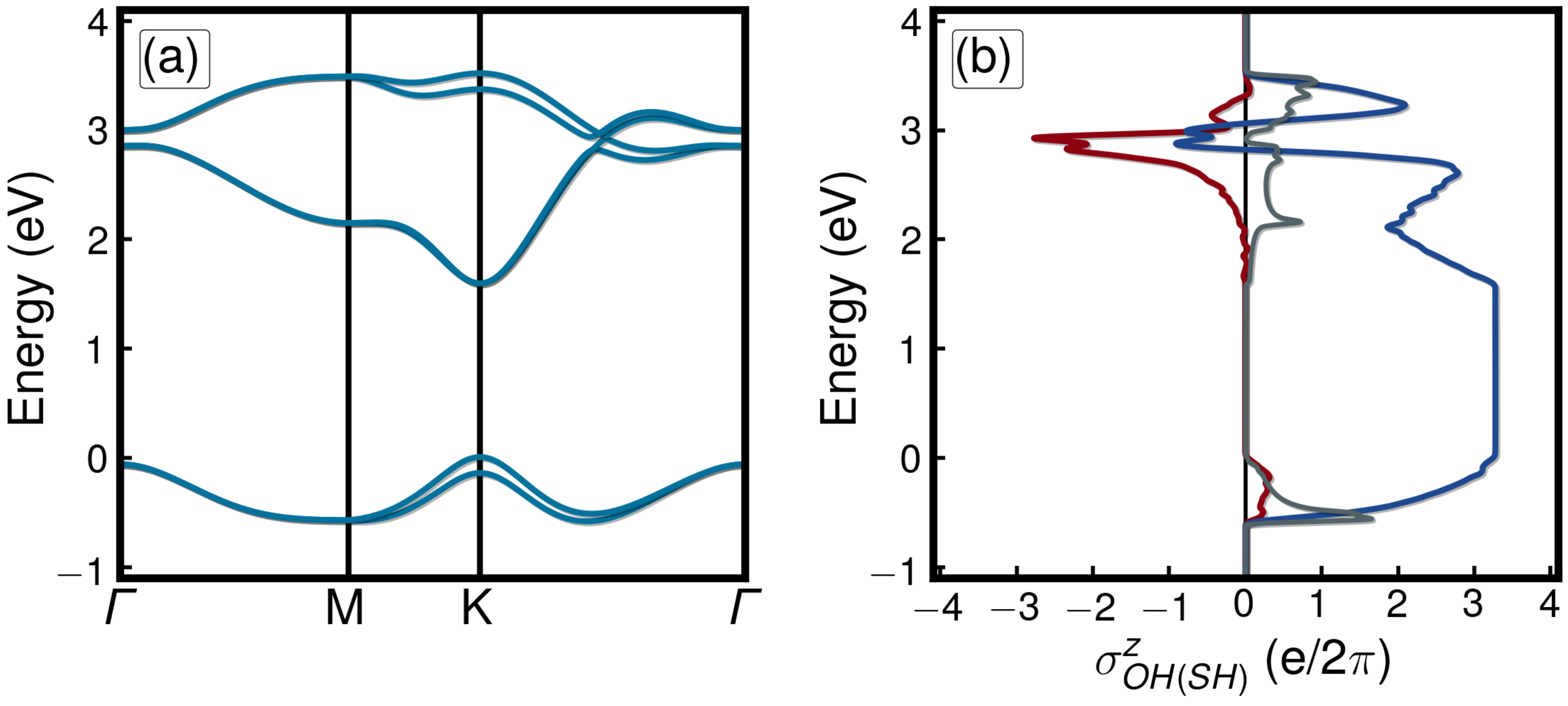}
	\includegraphics[width=0.95\linewidth,clip]{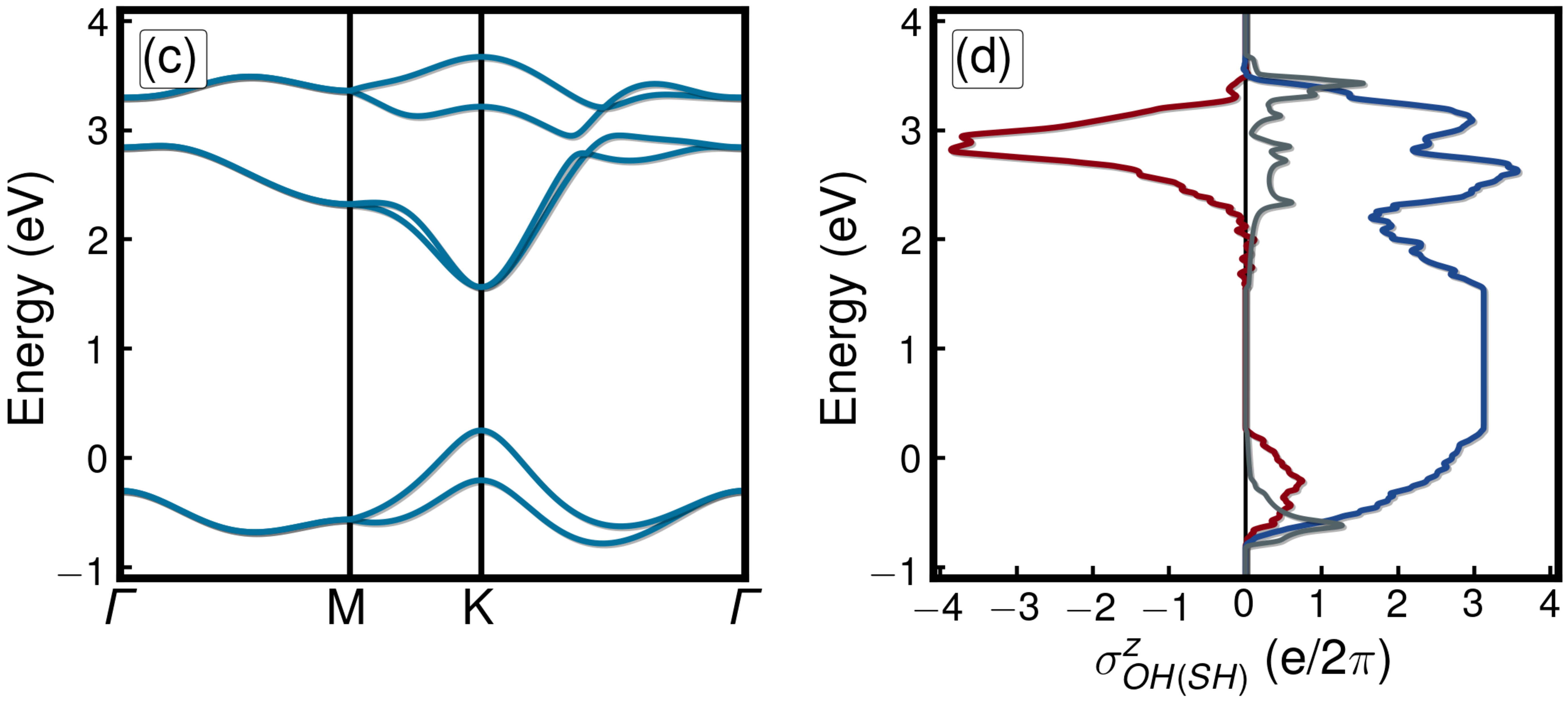}

	\caption{Band structures of MoS$_2$ (a) and WSe$_2$ (c) monolayers calculated along some high-symmetry directions in the 2D Brillouin zone (BZ) using the 3-bands TB model with SOC. The spin splitting in the valence band is $0.148$ eV for MoS$_2$ and $0.466$ eV for WSe$_2$. Panels (b) and (d) show the spin-Hall (red) and the orbital-Hall (blue) conductivities, together with the density of states in arbitrary units (grey), calculated as a functions of the Fermi energy for MoS$_2$ (b) and WSe$_2$ (d). The TB parameters were taken from Ref. \cite{ThreebandXiao}.}
	\label{fig:fig1} 
\end{figure}

\begin{figure}[h]
	\centering
	 \includegraphics[width=0.93\linewidth,clip]{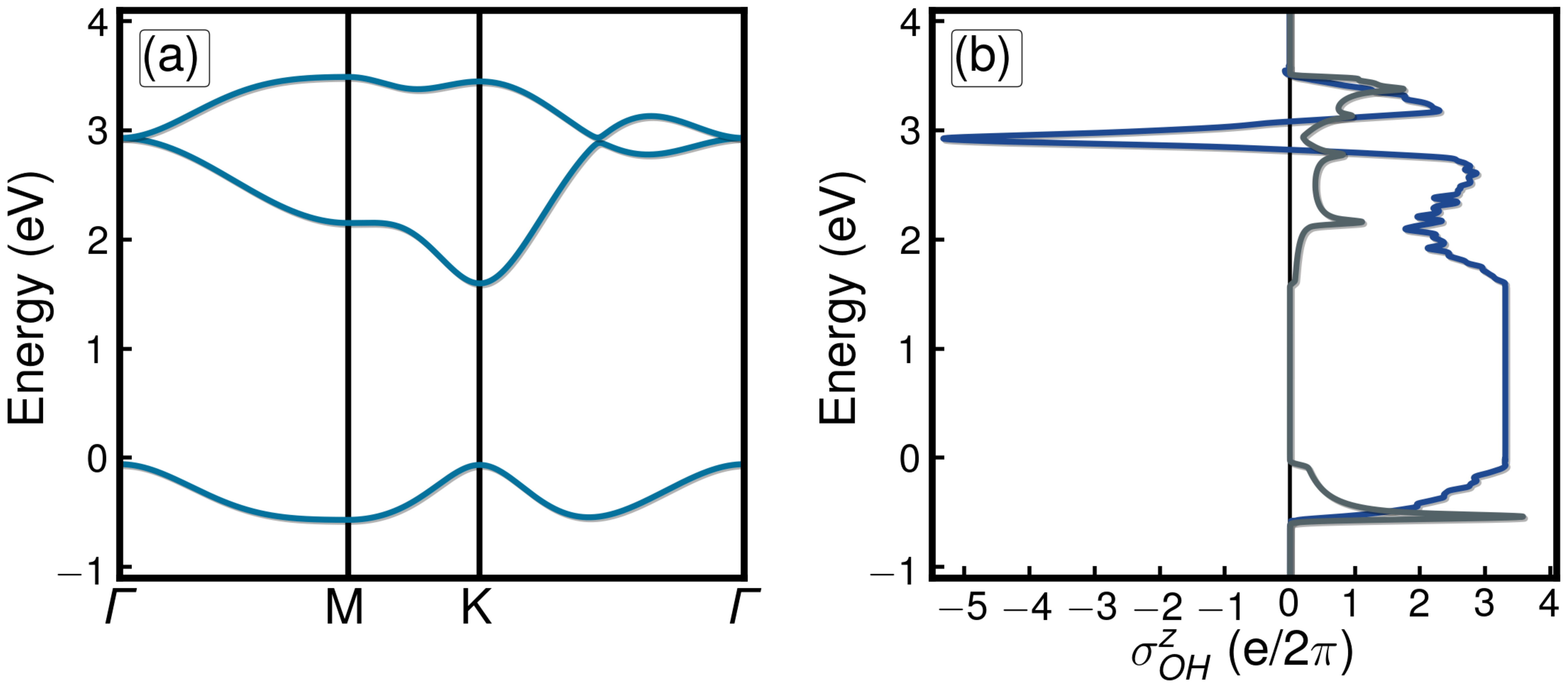}   
        \includegraphics[width=0.99\linewidth,clip]{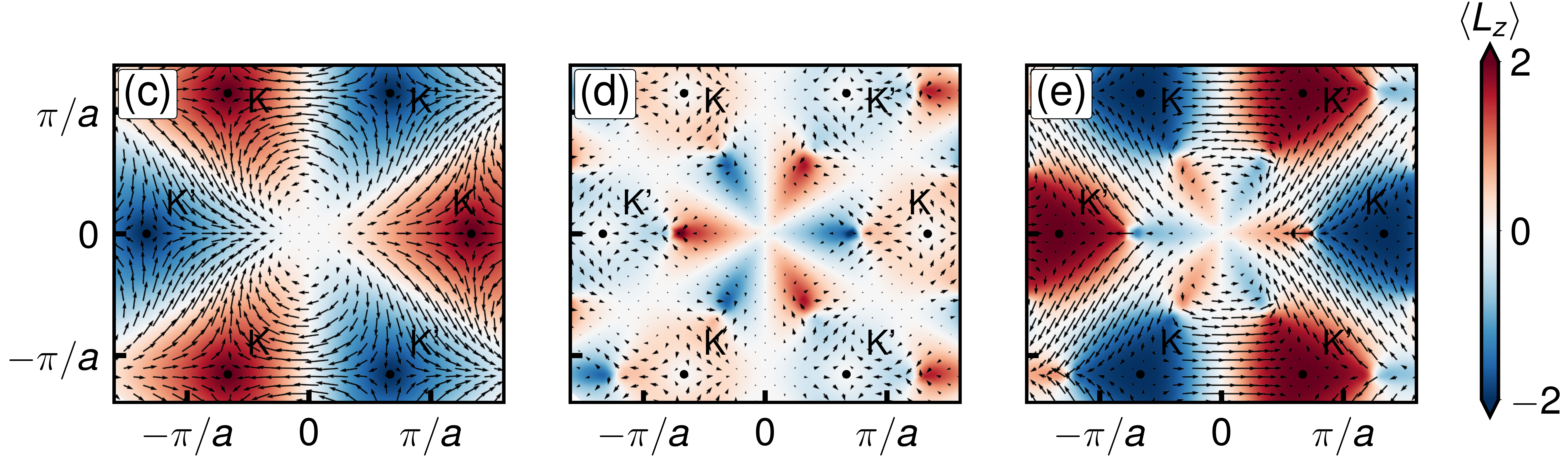}          
        \includegraphics[width=0.99\linewidth,clip]{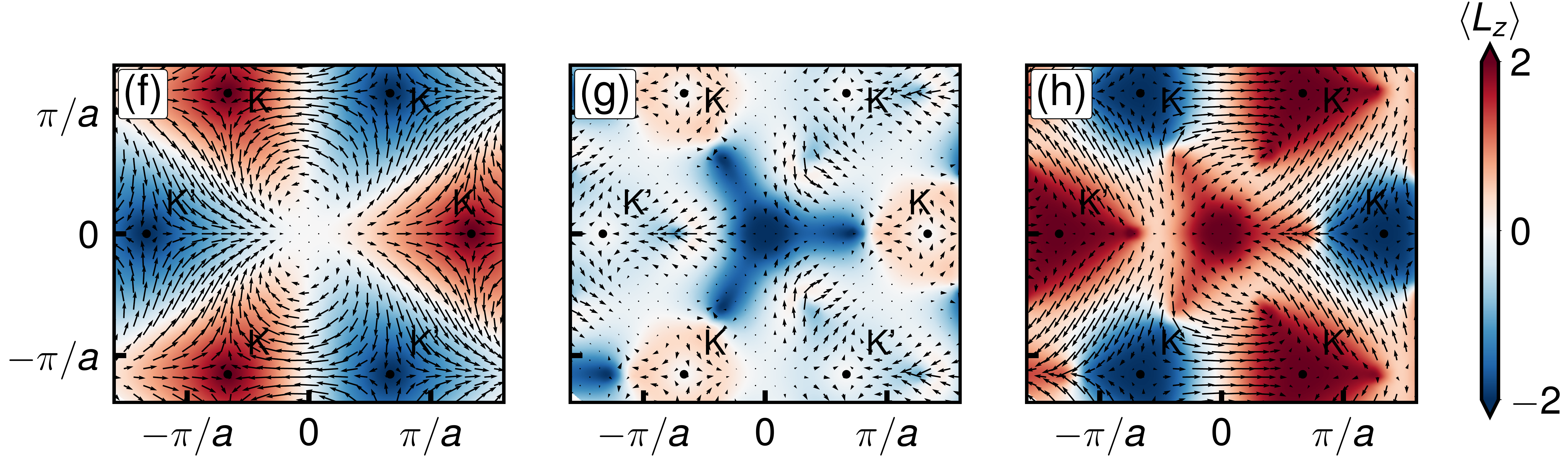}
	\includegraphics[width=0.99\linewidth,clip]{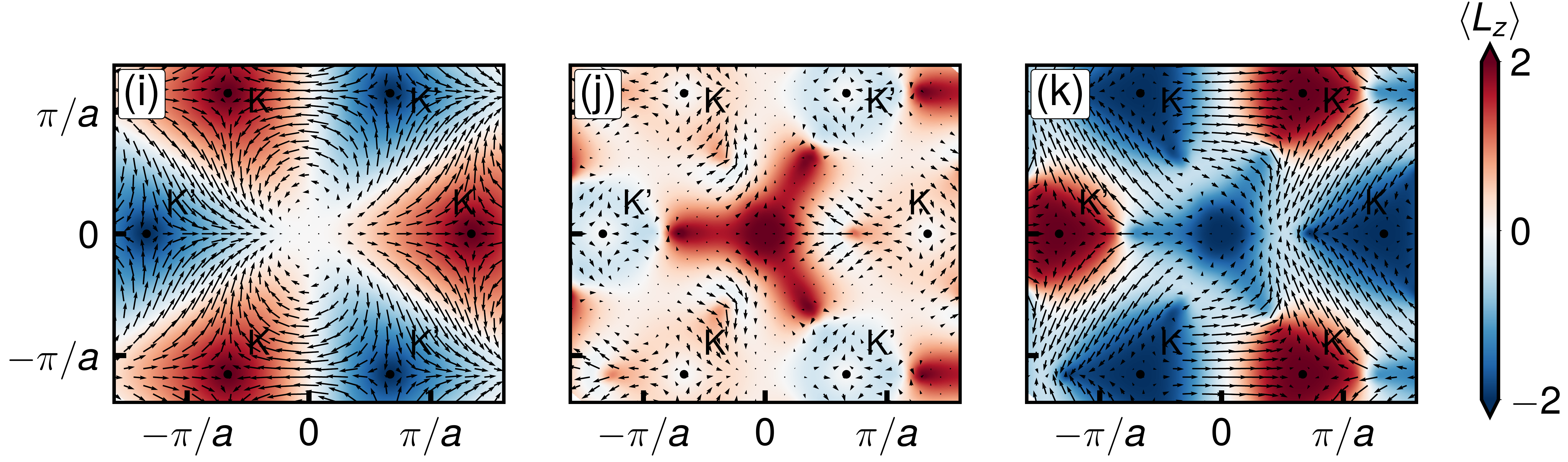}  
	\caption{(a) Band structure of a MoS$_2$ monolayer calculated along some high-symmetry directions in the 2D BZ using the 3-bands TB model without SOC. (b) and orbital-Hall (blue) conductivities, together with the density of states (grey), calculated as functions of the Fermi energy. The corresponding orbital-textures calculated for the valence band (c), conduction band (d), and highest energy band (e). Orbital textures calculated with SOC for the same sequence of bands: panels (f), (g) and (h) represent the $\uparrow$-spin sector; panels (i), (j) and (k) correspond to the $\downarrow$-spin sector}
	\label{fig:fig2}
\end{figure}

Panels (a) and (c) of Figure \ref{fig:fig1} show the band structures of MoS$_2$ and WSe$_2$, respectively, both calculated with the 3-bands TB model in the presence of SOC. The SOC causes a splitting in the valence band in the vicinity of the $K$ symmetry point, which is clearly more pronounced for WSe$_2$. Panels (b) and (d) exhibit the corresponding spin-Hall, and orbital-Hall conductivities, as well as the densities of states, calculated for MoS$_2$ and WSe$_2$, respectively. As expected, the spin-Hall conductivities for the H-TMDs with SOC vanish in the main energy gap because they are topologically trivial~\cite{tmdSHE}. However, the orbital-Hall conductivities are finite and exhibit plateaus of rather large magnitude within this energy range. We notice that these plateaus have similar values for MoS$_2$ and WSe$_2$, despite the markedly difference in their SOC intensities. In fact, the TMDs display very similar OHE plateaus even in the absence of SOC, as Fig. \ref{fig:fig2} illustrates. Panel (a) of Fig. \ref{fig:fig2} shows the band structure of MoS$_2$ calculated using the 3-bands TB model without SOC, and panel (b) depicts the corresponding orbital-Hall conductivity and the density of states.The spin-Hall conductivity in this case is zero and is worthless displaying it. Comparing Figs. \ref{fig:fig1} (b) and \ref{fig:fig2} (b), we clearly see that the presence of SOC in the TMDs affects the metallic phase, but just a little the orbital-Hall conductivity plateau. It slightly changes the plateau width by introducing a spin splitting in the valence band around the $K$ symmetry point, but the plateau hight is not altered. As we shall subsequently see, this comes from the fact that the accumulated in-plane orbital texture of the occupied states up to top of the valence band remains the same with the introduction of the SOC. One should notice that the OHE is a consequence of the existence of non-trivial angular-momentum-weighted (non-abelian) Berry curvature, as defined in references \cite{Mele_PRL_2019, Go-OHETexture, Murakami_NonabelianBerry}. This curvature can be non-trivial even in the presence of time-reversal and inversion symmetry and can be used as an alternative approach to calculate the OHE.

To get a further insight on the origin of this novel effect, it is instructive to enquire into the nature of the orbital textures in the TMDs, since they are linked to the OHE in three dimensional metals~\cite{Go-OHETexture} as well as in two-dimensional metals and insulators~\cite{us}. This simplified 3-bands TB model is restricted to a sector of the $L=2$ angular momentum vector space spanned only by the eigenstates of $L_z$: $d_{z^2}$ and $(d_{xy}\pm id_{x^2-y^2})/\sqrt{2}$ associated with $m_l=0, \pm2 \hbar$, respectively. Within this sector, it is useful to introduce a pseudo angular momentum $SU(3)$-algebra. The angular momentum operator components $\mathbf{L}^{x,y}$ in this basis can be obtained from $\mathbf{L}^{z}$, imposing that $\big[\mathbf{L}^{\alpha}, \mathbf{L}^{\beta}\big]=i\epsilon_{\alpha \beta \delta} \mathbf{L}^{\delta} $. We can also define the orbital texture associated with each electronic band, 
\begin{eqnarray}
\vec{\mathbf{L}}_{n,s}(\vec{k})=\sum_{\mu=x,y,z} \big<\psi_{n,s}^{\vec{k}}\big| \mathbf{L}^{\mu}\big|\psi_{n,s}^{\vec{k}}\big> \hat{e}_{\mu},
\label{Orbtexture}
\end{eqnarray}
where $\hat{e}_{x,y,z}$ are the Cartesian unit vectors, $|\psi_{n,s}^{\vec{k}}\big>$ represent the eigenstates associated with the energy bands $E_{n,s}^{\vec{k}}$, where $n=1,2,3$ label the 3-bands in increasing order of energy and s designate the spin sector.

Let us begin by analysing the orbital angular momentum textures for MoS$_2$ in the absence of spin-orbit coupling. Panels (c), (d) and (e) of Fig.\ref{fig:fig2} illustrates both the out-of-plane (color map) and the in-plane (vector field) projections of the orbital textures, calculated without SOC, for the energy bands $n=1,2,3$, respectively.
Panel (c) evinces that the valence bands of TMD monolayers exhibit an orbital-valley locking, showing that this effect precedes the well stablished spin-valley locking in TMDs and takes place even in the absence of SOC. In panel (d) we see that the out-of-plane component for the $n=2$ conduction band vanishes at the valleys. The orbital textures of the valence and conduction bands near $K$ and $K'$ as well as the orbital-valley locking are consistent with experimental optical characterisation of  the valley-Zeeman effect in TMD monolayers ~\cite{ValleyZeemanExitons}. Panel (e) also shows orbital-valley locking even in absence of SOC for the highest energy band. 
 
The OHE is caused by the dynamics of the in-plane orbital texture under the influence of an external longitudinal electric field, similarly to the SHE in the presence of Rashba SOC \cite{Sinova_PRL_2004}. It is clear from panels (c), (d) and (e) of Fig. \ref{fig:fig2} that the in-plane component of the orbital texture is stronger near the $K$ ($K'$) points and thus, the main contribution for the OHE in the insulating phase comes from the valley states where the simplified 3-bands model works pretty well. The contribution from the rest of the BZ affects the OHE quantitatively only. Fig. \ref{fig:fig2} (b) clearly shows that the OHE exists within the electronic energy band gap even in the absence of SOC. We also have checked that the inclusion of up to three next nearest neighbours hopping integrals in our 3-bands TB model \cite{ThreebandXiao} has a negligible effect on the orbital texture and in the OHE plateau, although it modifies OHE in the metallic phase \cite{SuplementaryMaterial}. The in-plane texture of the valence band near valleys is similar to a Dresselhaus orbital texture found for the $p_x$-$p_y$ model in honeycomb lattice, which also exhibits OHE in an insulating phase \cite{us}. When SOC is included  the orbital-valley locking persists, as panels (f)-(k) of Fig. \ref{fig:fig2} show. The in-plane orbital texture is not qualitatively affected by the SOC, but the out-of-plane texture is strongly modified near the $\Gamma$ point for the conduction band. In its vicinity we find full out-of-plane orbital polarisations with reverse directions for opposite spin sectors. At the $K$ and $K'$ valleys, the out-of-plane orbital texture is again only quantitatively influenced by the SOC, and we see that it has the same signal independent of spin sector. Similarly to what we have previously pointed out, it is the in-plane component of the orbital textures displayed by panels (f)-(k) of Fig. \ref{fig:fig2} that generates the OHE illustrated in Fig. \ref{fig:fig1} (b) for MoS$_2$ \cite{us}. 

It is worth mentioning that vertex corrections are not expected to suppress the OHE in the TMDs because of their symmetry \cite{Dimitrova-vertex}, and should not affect the OHE plateau due to the absence of Fermi-surface in insulators \cite{Milletari-vertex}. We also remark that OAM in the insulator phase of the TMDs can be partially transported by well-known edge-states of TMDs in zigzag nanoribbons \cite{IOPEdgeMoS2}, but additional contributions may come from the bulk \cite{Mele_PRL_2019}.

\begin{figure}[h]
	\centering
	 \includegraphics[width=0.99\linewidth,clip]{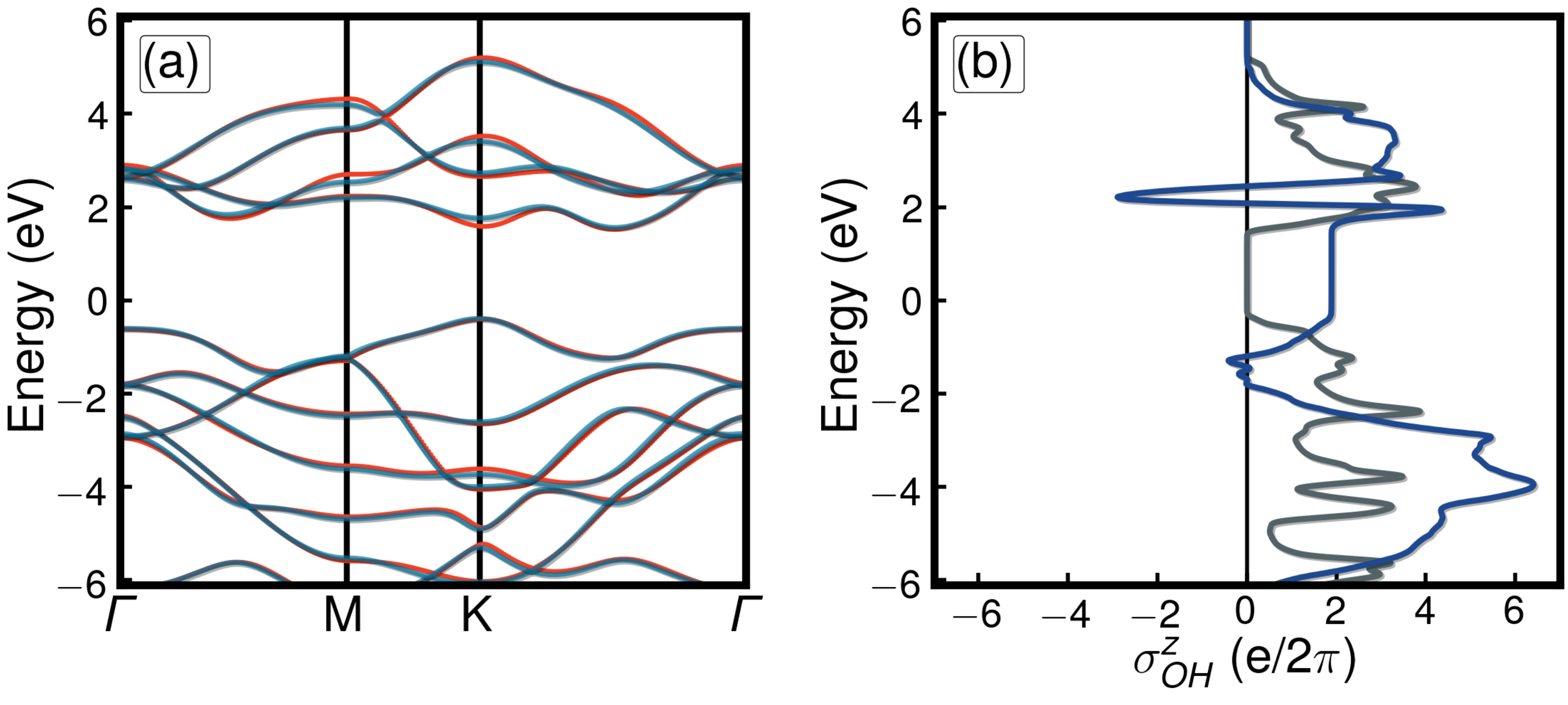} 
	 \includegraphics[width=0.99\linewidth,clip]{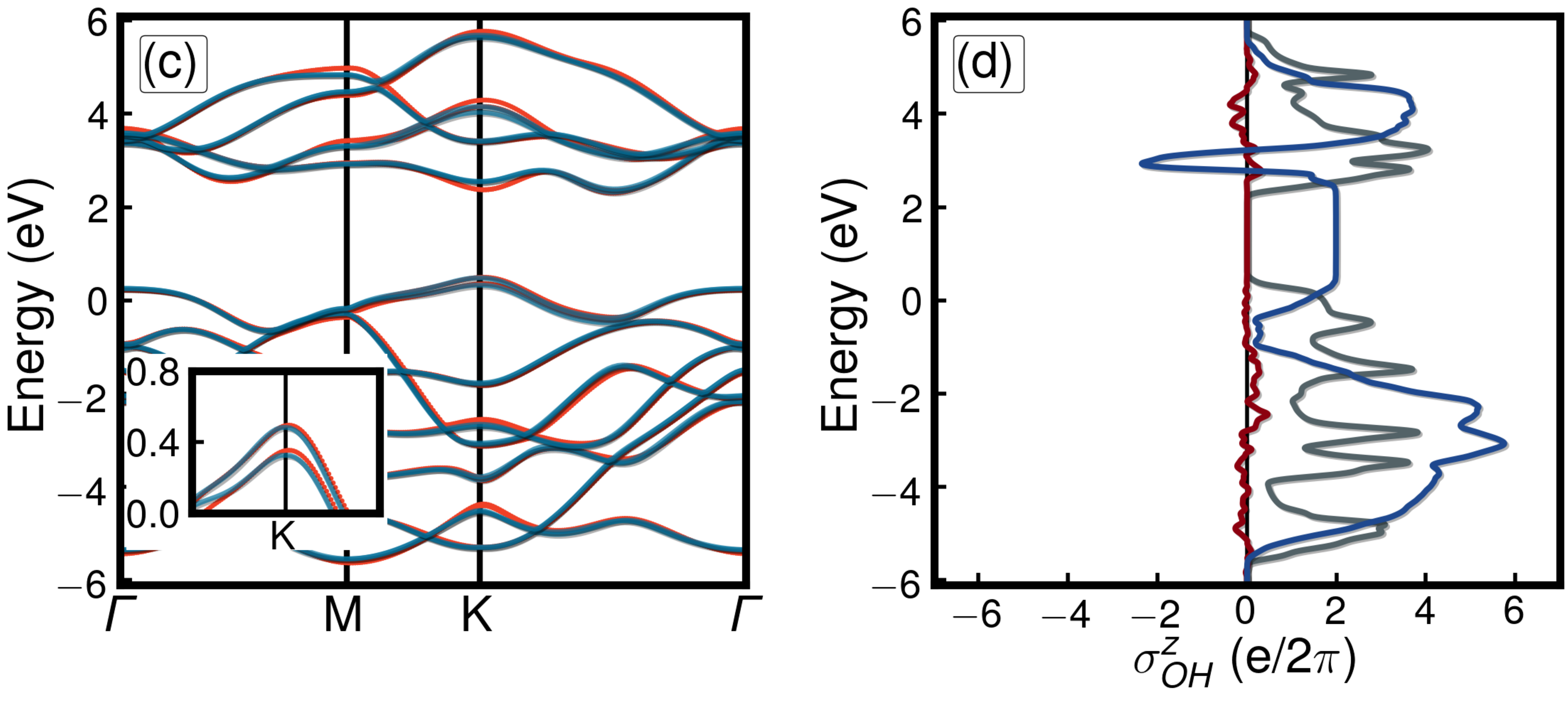}         
        \caption{(a) Comparison between the band structures of a MoS$_2$ monolayer calculated without SOC employing DFT (blue) and the effective TB model that consider all  hopping integrals with  energies higher than 0.0125 $|t|_{max}$ (red); (b) Density of states (grey) and OH conductivity (blue) calculated without SOC using the effective TB model. Panels (c) and (d) illustrate the same as (a) and (b), but with SOC. The inset in panel (c) highlights the spin-splitting of the valence band near the $K$ point which is $\approx 0.148$ eV.}
	\label{fig:fig3}
\end{figure}

So far we have used a simplified 3-bands model for unveiling the main features of the OH {\it insulator} phase of the TMDs. This model describes very well the physics near valleys, which are responsible for the main contribution to the OHE plateau, and give us good insights about the phenomena. Still, it is crucial to verify if our findings are endorsed by a more realistic calculation that is not restricted to the vicinity of the $K$ point. This is especially relevant in this scenario because in a real material other atomic orbitals could contribute to the transport properties of these systems. It is instructive, for example, to inquire into the contributions to the OHE coming from the atomic orbitals of the chalcogen atoms. To this end, we have performed density-functional theory (DFT) calculations, within the local-density approximation (LDA), using the Quantum Espresso numerical packages \cite{Giannozzi2009}. We have employed norm-conserving and fully relativistic pseudopotentials \cite{Hamann2013,VanSetten2018}, and generated a basis of atomic orbitals using Wannier90 \cite{Pizzi2019} for MoS$_2$, that includes the $s$, $p_x$, $p_y$ and $p_z$ orbitals of the chalcogen  atoms and the five $d$ orbitals of the TM atoms (see exact composition of states at $K$ in Ref. \cite{MolinaSanchez2015}). The effective tight-binding Hamiltonian obtained with Wannier90 is then exported to KITE, with the help of PythTB scripts~\cite{vanderbilt_2018} to carry out our quantum transport calculations. Panels (a) and (c) of Fig. \ref{fig:fig3} illustrate comparisons between band structure calculations for MoS$_2$ obtained by DFT and by the effective TB Hamiltonian, without and with SOC respectively. We have considered all hopping integrals with energies higher than $0.0125$ of the maximum hopping value $|t|_{max}$. The agreement between the two methods is excellent, especially in the vicinity of the band gap.

In panel \ref{fig:fig3} (b) we present the density of states and OH  conductivity calculated with the effective Hamiltonian taking into account 13 orbitals per unit cell, where we see that a sizeable OHE  within the MoS$_2$ electronic gap. When SOC is taken into account, a new parametrised Hamiltonian is obtained and we see that it also reproduces the corresponding DFT band structure quite well, including the spin-splitting of the valence band in the vicinity of the Dirac pints - see the inset in panel \ref{fig:fig3} (b).  Figure \ref{fig:fig3} (d) depict the density of states, SH and OH conductivities calculated with SOC for MoS$_2$. It clearly shows that the OHE is much stronger than the SHE for H-TMDs. The OH plateau height calculated with this full TB parametrisation is approximately 30$\%$ smaller than the one obtained with the simplified 3-bands TB model. This seems reasonable because the OHE  depends upon details of the electronic structure~\cite{us} and the OH plateau height increases with the $\mathbf{L}^z$ projected Berry curvature value. Since the curvature of the band structure in the vicinity of the Dirac point is more pronounced for the 3-bands model, it is likely to yield a higher plateau. We also note that SOC clearly does not affect much the OHE within the energy gap for the H-TMDs, reinforcing the fact that the OHE is essentially linked to the orbital composition and symmetry of the system.

Regarding the observation of the OH insulating phase, we recall that within the energy gap of the H-TMDs no spin Hall current is generated by a longitudinally applied electric field, but a pure orbital Hall current is produced. In order to detect it we envisage this pure orbital angular momentum current being injected into a suitable system with non negligible SOC, thereby  inducing a spin current that can be detected by the inverse SHE in a non-magnetic material or by means of exerted torques on a ferromagnet, as suggested in references \cite{Orbital-torque} and \cite{Zheng_et_al}.

It is noteworthy that the valley Hall effect (VHE) displayed by TMDs bears similarities with the OHE discussed here, but they are not the same phenomena. The VHE is associated with the valley OAM \cite{LightMatTMD, Berry_RMP_Niu_2010} and requires no sizeable orbital angular momentum texture to occur. It may happen, for example, in graphene when sub-lattice symmetry is broken ~\cite{qniurev}. In contrast, the OHE is related to the existence of orbital angular momentum textures in reciprocal space, which arise from characteristics of the $p$ and $d$ atomic orbitals and lattice symmetry. It does not rely on the existence of well-defined valleys and takes place also in systems with indirect band gaps~\cite{us}. When the VHE contribution is written in terms of its OAM~\cite{Xiao2007}, the total OHE is a sum of the two contributions. Presently, however, it may be experimentally  challenging to discriminate them in the TMDs. Careful analysis of their roles in non-local transport experiments \cite{non-loc-measurement} and valley-selective dichroism \cite{Xiao_PRL_2012, TCao_NatureComm} may possibly lead to progress in this direction, as well as further investigations on orbital and valley angular momentum coherence lengths.


In summary, with quantum transport calculations we have shown that TMDs such as MoS$_2$ and WSe$_2$ are orbital Hall insulators and can host sizeable OHE for energies within their electronic energy gaps. The use of OAM as an information carrier in TMDs widens the development possibilities for novel spin-orbitronics two-dimensional devices. 

\begin{acknowledgments}
	We acknowledge CNPq/Brazil, CAPES/Brazil, FAPERJ/Brazil and INCT Nanocarbono for financial support, and NACAD/UFRJ for providing high-performance computing facilities.  TGR acknowledges COMPETE2020, PORTUGAL2020, FEDER and the Portuguese Foundation for Science and Technology (FCT) through project POCI-01- 0145-FEDER-028114. TPC acknowledges S\~ao Paulo Research Foundation (FAPESP) grant 2019/17345-7
	
\end{acknowledgments}
\begin{widetext}
\begin{center}
{\bf Supplementary material for ``Orbital Hall Insulating Phase in Transition Metal Dichalcogenide Monolayers''}	
\end{center}

\section{Kubo formula for linear response conductivity} 
In the main text, we used the Kubo-Bastin formula to compute spin and orbital Hall conductivities for tight-binding models in real-space. Here we shall briefly examine an alternative formulation for calculating orbital and spin Hall conductivities, which is equivalent to Eq. 2 of the main text. With this approach the spin Hall (SH) and orbital Hall (OH) $\eta$-polarized response, in $\hat{y}$ direction, to an electric field applied in $\hat{x}$ direction is given by,

\begin{eqnarray}
	\sigma^{\eta}_{OH(SH)}=\frac{e}{\hbar} \sum_{n\neq m}\sum_{s=\uparrow,\downarrow} \int_{B.Z.} \frac{d^2k}{(2\pi)^2} (f_{m\vec{k}}-f_{n\vec{k}}) \Omega_{n,m,\vec{k},s}^{X_{\eta}}, \label{Kubo1}
\end{eqnarray}
\begin{eqnarray}
	\Omega_{n,m,\vec{k},s}^{X_{\eta}}=\hbar^2 \text{Im} \Bigg[ \frac{\big<\psi^s_{n,\vec{k}}\big|j_{y}^{X_{\eta}}(\vec{k})\big|\psi^s_{m,\vec{k}}\big>\big<\psi^s_{m,\vec{k}}\big|v_x(\vec{k})\big|\psi^s_{n,\vec{k}}\big>}{(E^s_{n,\vec{k}}-E^s_{m,\vec{k}}+i0^+)^2}\Bigg], \label{Kubo2}
\end{eqnarray}
where $\sigma^{\eta}_{OH(SH)}$ is the orbital Hall (spin Hall) DC conductivity with polarizsation in $\eta$-direction, $\Omega_{n,m,\vec{k},s}^{X_{\eta}}$ is the  gauge-invariant spin and orbital weighted Berry curvatures. In Eq. \ref{Kubo2}, $E^s_{n(m),\vec{k}}$ and $|\psi^s_{n(m),\vec{k}}\big>$ are eigenvalues and eigenvectors of the tight-binding Hamiltonian in the reciprocal space, for $n(m)$ Bloch band, and $s=\uparrow,\downarrow$ labels the spin-sector. The velocity operators are defined as $v_{x(y)}(\vec{k})=\partial H(\vec{k})/\partial \hbar k_{x(y)}$, where $H(\vec{k})$ is the tight-binding Hamiltonian in reciprocal space. The current density operator component  in $\hat{y}$ direction is defined as $j_y^{X_{\eta}}(\vec{k})=\big(X_{\eta}v_y(\vec{k})+v_y(\vec{k})X_{\eta} \big)/2$, where $X_{\eta}=\hat{\ell}_{\eta}(\hat{s}_{\eta})$ for OH (SH) conductivities polarized in $\eta$ direction.

\section{OHE for 3-bands model using Kubo Formula for MoS$_2$}

We have examined the orbital-Hall insulating phase of TMDs by considering a simplified 3-bands model that includes only three atomic $d$ orbitals ($d_{z^2}, d_{xy}, d_{x^2+y^2}$) of the transition metal (TM) atoms arranged in a triangular lattice.  The hopping integrals for his effective 3-bands model are computed by assuming the $D_{3h}$-point group symmetry of the TMD in the H structural phase, which captures the effect of the hybridization of the $d$-orbitals of TM with the $p$-orbitals of the chalcogenes. This effective model was developed by Liu et. al. \cite{ThreebandXiao}. As a first approximation, they consider first nearest neighbour hopping integrals only in the triangular lattice. This simplified model was then used to fit the energy spectrum of DFT calculations near the K-points  (valleys) of the two dimensional (2D) first Brillouin Zone (BZ). The TB parameters obtained for this simplified model and the corresponding Hamiltonian written in the reciprocal space are registered in reference [\onlinecite{ThreebandXiao}]. 

The representation of the $z$-component of the orbital angular momentum operator in this simplified 3-bands model is given by
\begin{eqnarray}
\mathbf{L}^{z}= \hbar\begin{bmatrix}
0 & 0 & 0\\
0 & 0 & 2i\\
0 & -2i & 0
\end{bmatrix}.
\label{lz}
\end{eqnarray}

We have mentioned in the main text that the Kubo-Bastin formula used in our primary calculations is equivalent to Eqs. (\ref{Kubo1}) and (\ref{Kubo2}) shown above. In order to illustrate such equivalence we compare results obtained for the orbital Hall conductivity of MoS$_2$ with both approaches. They are depicted in Fig. \ref{fig:compareplot}, where  the blue solid lines represent the results calculated with the the Kubo-Bastin formula and the black dashed lines the ones obtained with Eqs. (\ref{Kubo1}) and (\ref{Kubo2}). Panels (a) and (b) show the calculated results without and with SOC, respectively. The agreement between the two approaches is excellent, as expected. 
 
\begin{figure}[h]
	\centering
	\includegraphics[width=0.4\linewidth]{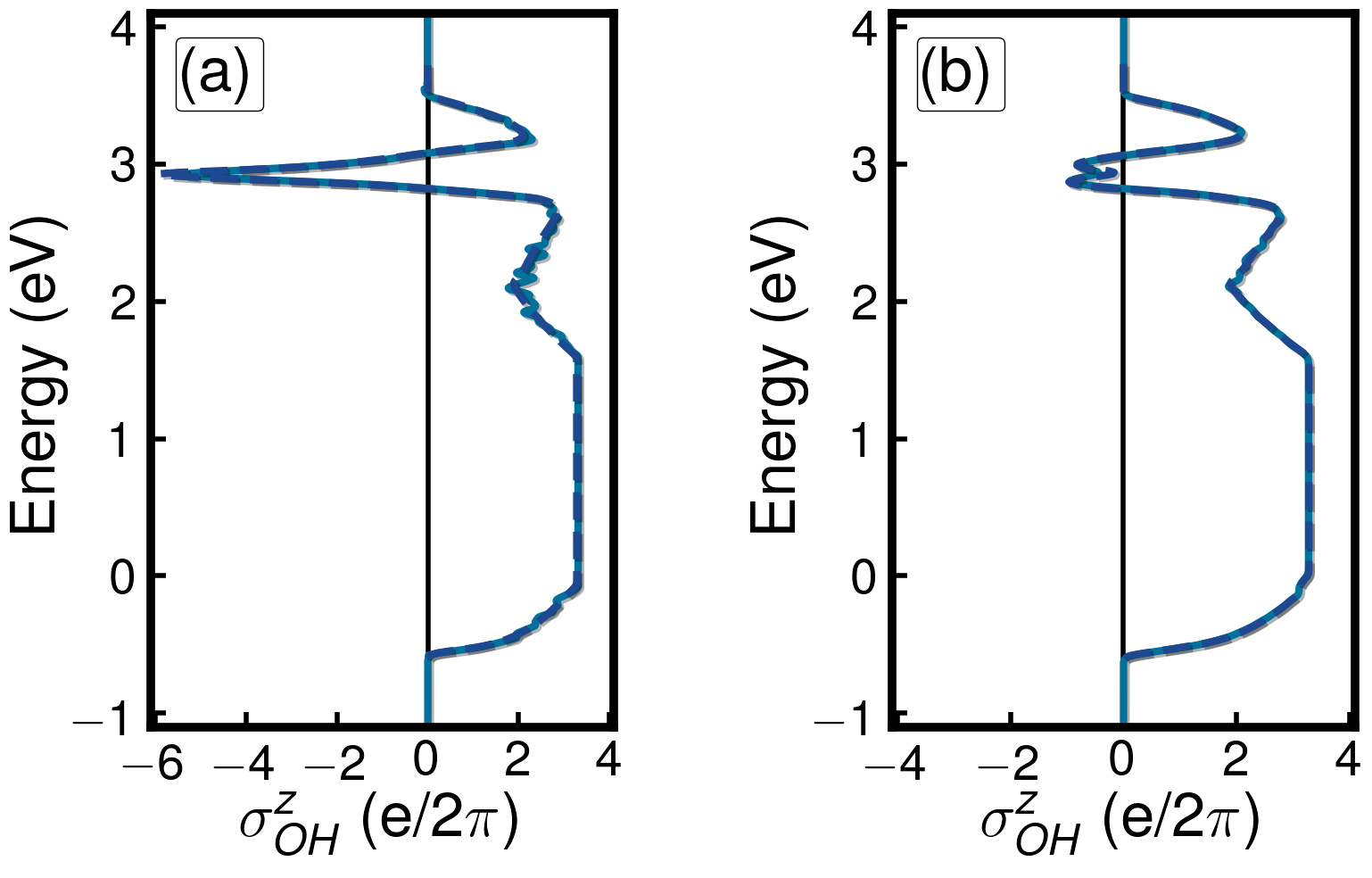}
	\caption{Comparison between the orbital Hall conductivities calculated for MoS$_2$ using the 3-bands tight-binding model with first nearest neighbours only, employing the Kubo-Bastin (blue solid line) and the Kubo formulas (black dashed line). Panels (a) and (b) illustrate results obtained without and with SOC, respectively.}
	\label{fig:compareplot}
\end{figure}

As reported in in reference [\onlinecite{ThreebandXiao}], the TB parameterization with first nearest-neighbour (1nn) hopping integrals only works well in the vicinity of the $K$-points, but relatively far from them the energy spectrum is not well reproduced. An improvement can be obtained by including up to three nearest-neighbor (3nn) hopping integrals in this model, reproducing very well the band edges, higher energy valence-band ($n=1$) and lower energy conduction-band ($n=2$) in the vicinity of the semiconducting gap obtained by DFT calculations. The detailed construction and the parameters obtained with both first 1nn and up to 3nn approximations can be found also in same reference.

In our work, we have presented results for the OH conductivity calculated with 3-bands TB model taking into account 1nn-hopping integrals only, and mentioned that it gives a good description of the orbital-Hall plateau. Here, we compare results for the OHE calculated for MoS$_2$ with the 3-bands model considering 1nn hopping integrals and up to 3nn. To this end, we used the Kubo-formula given by the Eqs. (\ref{Kubo1}) and (\ref{Kubo2}) of this Supplementary Material (SM)  to compute the orbital Hall conductivities employing both approximations. 

Figure \ref{FIGSM1} shows the results for OHE obtained with the 1nn (blue dashed curves) and 3nn (black solid curves) approximations, in the presence (right panel) and in the absence (left panel) of spin-orbit coupling (SOC). It is worth noticing that the OH plateau height changes very little with the inclusion 3nn hopping integrals. This corroborates our reasoning that the OH insulator phase in TMDs is dominated by a Dresselhaus-like orbital-texture [\onlinecite{us}] near valleys, as mentioned in the main text. The metallic phase, however, is strongly affected by the inclusion of 3nn hopping integrals, as expected. The height of the OH plateau is proportional to the integral of the orbital Berry-curvature in the 2D first BZ. Figure \ref{FIGSM2} illustrates a density-plot of the orbital Berry-curvature for the valence band $n=1$, namely $\Omega_O (\vec{k})=\sum_{m=2,3}\Omega_{n=1,m}^{L_{z}}(\vec{k})$, in entire 2D first BZ for MoS$_2$. The results are obtained using the 3-bands model without SOC, taking into account 1nn hopping integrals only (the left panel) and up to 3nn (right panel). Although the orbital-Berry curvatures calculated with 1nn and up to 3nn are very different far from the valleys, the peaks near K-points are very similar - see color code in the figure (\ref{FIGSM2}) - leading to almost identical OH plateau heights in both cases. \\

\begin{figure}[h!]
	\includegraphics[width=0.85\linewidth,clip]{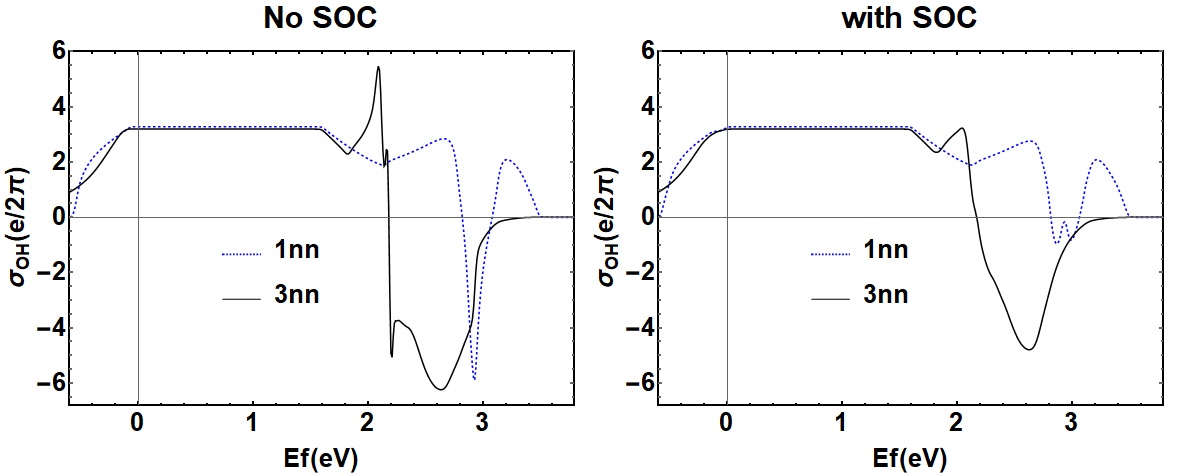}
	\caption{Orbital Hall conductivities computed using the Kubo formula for MoS$_2$ with (right) and without (left) SOC, for 3-bands model \cite{ThreebandXiao}. Blue dashed curves are the results for the 3-bands model using the first-nearest neighbor (1nn) hopping approximation, and black solid curves illustrate the results using {\it up to} third-nearest neighbor (3nn) hopping integrals \cite{ThreebandXiao}.}
	\label{FIGSM1} 
\end{figure} 

\begin{figure}[h!]
	\includegraphics[width=0.85\linewidth,clip]{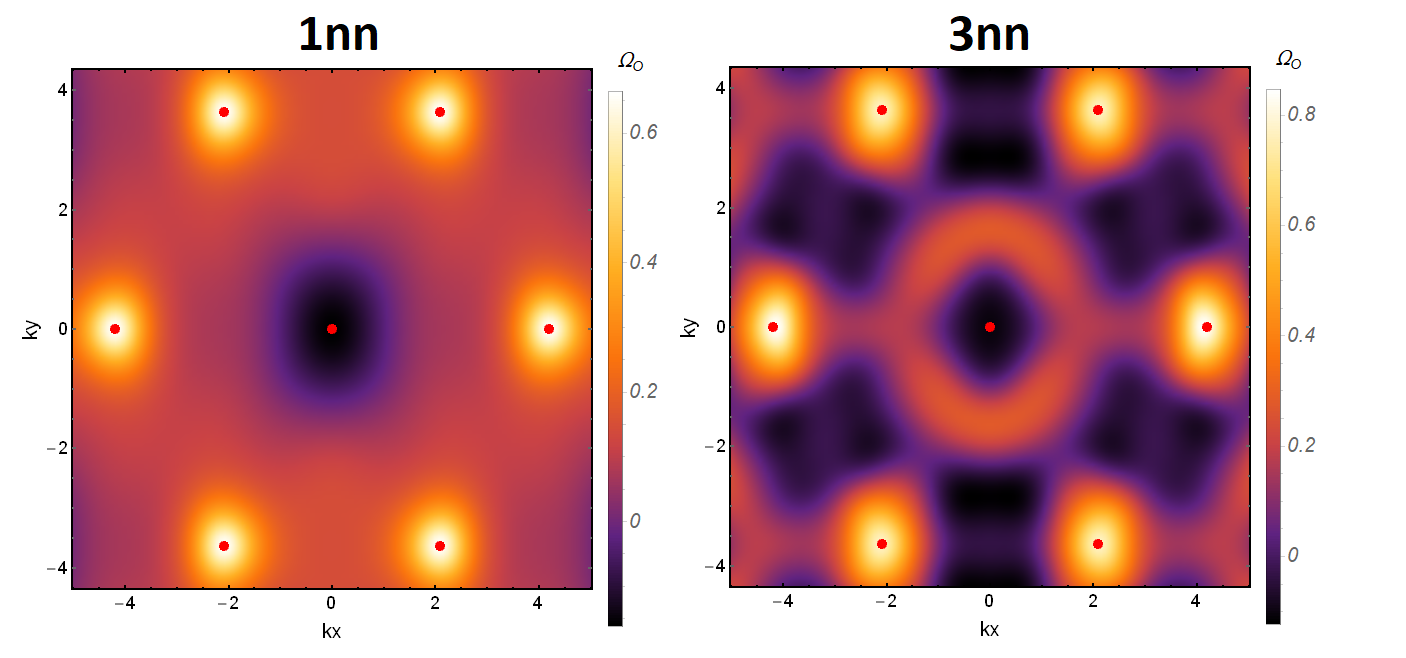}
	\caption{Density plot for the Orbital Berry-curvatures (Eq. \ref{Kubo2}) of the valence band of MoS2 calculated with 3-bands model, without SOC, employing 1nn (left panel) and up 3nn (right panel) hopping integrals \cite{ThreebandXiao}. The inclusion of SOC slightly modifies the weight of the Berry curvatures near valleys. {\it Red dots} at the hexagon vertices indicate the $K$-points (valleys), and the red dot in the center of the BZ highlight the $\Gamma$-point. We have used $\hbar=1$ in this color map.}
	\label{FIGSM2}
\end{figure}

\end{widetext}

\end{document}